\documentclass[journal]{IEEEtran}
\usepackage[utf8]{inputenc}
\usepackage[caption = false]{subfig}
\usepackage{comment}
\usepackage{cite}
\usepackage{amsmath,amsmath,bm,amsfonts}
\usepackage{graphicx}
\usepackage{textcomp}
\usepackage{booktabs}
\usepackage{todonotes}

\newcommand{\zvi}[1]{\mathbf{z}_{#1}}

\newcommand{\svi}[1]{\mathbf{s}_{#1}}

\title{Leveraging Pretrained Image-text Models for Improving Audio-Visual Learning}
\author{Saurabhchand Bhati$^{\dagger}$, Jes\'us Villalba$^{\dagger,\ddagger}$, Laureano Moro-Velazquez$^{\dagger}$, Thomas Thebaud$^{\dagger}$, Najim Dehak$^{\dagger,\ddagger}$ \\
$^{\dagger}$Center for Language and Speech Processing, Johns Hopkins University, USA \\
 $^{\ddagger}$Human Language Technology Center of Excellence, Johns Hopkins University, USA \\
 \{sbhati1,jvillalba,laureano,tthebau1,ndehak3\}@jhu.edu}
\date{November 2022}

\begin{document}

\maketitle

\begin{abstract}
Visually grounded speech systems learn from paired images and their spoken captions. Recently, there have been attempts to utilize the visually grounded models trained from images and their corresponding text captions, such as CLIP, to improve speech-based visually grounded models' performance. However, the majority of these models only utilize the pretrained image encoder. Cascaded SpeechCLIP attempted to 
generate localized word-level information and 
utilize both the pretrained image and text encoders. 
Despite using both, they noticed a substantial drop in retrieval performance. We proposed Segmental SpeechCLIP which used a hierarchical segmental speech encoder to generate sequences of word-like units. We used the pretrained CLIP text encoder on top of these word-like unit representations and showed significant improvements over the cascaded variant of SpeechCLIP. Segmental SpeechCLIP directly learns the word embeddings as input to the CLIP text encoder bypassing the vocabulary embeddings. Here, we explore mapping audio to CLIP vocabulary embeddings via regularization and quantization. As our objective is to distill semantic information into the speech encoders, we explore the usage of large unimodal pretrained language models as the text encoders. Our method enables us to bridge image and text encoders e.g. DINO and RoBERTa trained with uni-modal data. Finally, we extend our framework in audio-only settings where only pairs of semantically related audio are available. Experiments show that audio-only systems perform close to the audio-visual system. 
\end{abstract}

\section{Introduction}
Speech processing systems aided by large amounts of labeled data and computational resources achieve remarkable performance~\cite{baevski2020wav2vec,synnaeve2020end,wang2020transformer}.
However, vast amounts of labeled data are not available for most languages, and transcribing a large amount of speech data is expensive. Therefore, there has been a lot of interest in developing methods to learn useful information from unlabeled data~\cite{badino2014auto,lee2012nonparametric,siu2014unsupervised,kamper2017segmental,bhati2017unsupervised,kamper2017embedded,bhati2018phoneme}. Recently, self-supervised learning (SSL) methods have emerged as a significant paradigm for learning representations from unlabeled audio data~\cite{oord2018representation,schneider2019wav2vec,baevski2020wav2vec}. In SSL methods, the model is trained to solve a pretext task for which labels can be generated from the raw audio. Some common pretext tasks include masked language modeling~\cite{baevski2020wav2vec,chung2021w2v}, next frame prediction~\cite{oord2018representation}, next segment prediction~\cite{bhati21_interspeech,bhati2022unsupervised}, and masked reconstruction~\cite{liu2020mockingjay,liu2021tera}. Speech systems built on top of these SSL representations require much less labeled data to match the performance of systems built without them~\cite{baevski2020wav2vec}.  
Another direction is to use multimodal data and extract useful information to improve performance in a given modality.

Parallel text and image data have been leveraged for learning representations that help downstream performance in both modalities~\cite{chen2020uniter,radford2021learning}. Contrastive language image pretraining (CLIP) learns to align the parallel image and text data crawled from the internet~\cite{radford2021learning}. CLIP shows remarkable performance in zero-shot setting for image classification and image/text retrieval from text/images~\cite{radford2021learning}. 
\begin{figure}
    \centering
    \includegraphics[width=\columnwidth]{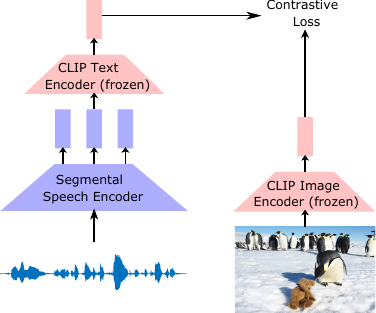}
    \caption{Overview of the proposed Segmental SpeechCLIP.}
    \label{fig:overview}
    \vspace{-2mm}
\end{figure}
Parallel images and spoken captions have also been leveraged to improve speech processing systems~\cite{harwath2018jointly,peng2022fast,hsu2019transfer,peng2022word,hsu2021text,peng2022fastvgsplus}. These systems are commonly referred to as visually grounded speech (VGS) systems. VGS systems have been shown to improve speech systems performance for speech recognition~\cite{hsu2019transfer}, word discovery~\cite{peng2022word}, and speech synthesis~\cite{hsu2021text}. VGS models trained with just retrieval loss can learn semantic~\cite{peng2022fastvgsplus} and word-level information, such as boundaries~\cite{peng2022word} from speech. 

Recently, there have been efforts to utilize CLIP for 
improving the performance of VGS systems. However, most of these systems only utilize the CLIP model's image encoder. WAV2CLIP~\cite{wu2022wav2clip} and the parallel variant of speechCLIP~\cite{speechclip2022} generate a single representation per utterance summarizing the information. This global representation is then used for classification and retrieval tasks. There are no constraints to localize word-level information. Guidance from models trained on text data, such as CLIP text encoder, could help extract semantic information from speech. For example, unsupervised ASR systems use nonparallel text data and a pronunciation lexicon for speech recognition in the absence of transcribed speech data. These were some of the motivations behind the cascaded variant of speechCLIP. 

The cascaded SpeechCLIP~\cite{speechclip2022} model appends K learnable CLS tokens to the utterance to extract the most important keywords. Vector quantization is then used to map these keywords to CLIP's subword embeddings. The frozen text encoder is used on top to generate sentence embedding. A frozen CLIP image encoder is used to extract image representations. However, the cascaded variant has significantly lower retrieval recall scores than the parallel variant.

In this work, we extend Segmental SpeechCLIP~\cite{bhati23_interspeech} to utilize text encoders trained with multimodal data such as image-text and unimodal data i.e. text or images. Segmental SpeechCLIP~\cite{bhati23_interspeech} improves the keywords extraction from speech utterances and better utilization of the text CLIP encoder. Segmental SpeechCLIP uses a segmental speech encoder based on Segmental Contrastive Predictive Coding(SCPC)~\cite{bhati21_interspeech,bhati2022unsupervised} to extract sequences of word-like units from audio. 
We stack a pretrained text encoder on top of the generating sub-words to extract a sentence embedding. A pretrained image encoder is used to extract image representations. The model tries to align the semantically related images and their spoken captions. Using the text encoder, we want to infuse semantic information in the speech encoder.
The overall architecture is shown in Fig.~\ref{fig:overview}. We show that our Segmental SpeechCLIP significantly outperforms the cascaded variant of SpeechCLIP. The cascaded speechCLIP model discovers a fixed number of keywords, i.e., eight from the utterances. Our Segmental SpeechCLIP automatically deduces the number of word-like units in an utterance. There is no implicit constraint to enforce the temporal structure on the discovered keywords in the cascaded speechCLIP, whereas Segmental SpeechCLIP, by design, discovers word-like units in temporal order. Instead of quantization and mapping the keywords to CLIP vocabulary embeddings, we directly learn the subword embeddings. 

We experiment with adding additional constraints on the model to map the learned embeddings to CLIP vocabulary embeddings. We explore regularization and quantization as constraints. Successfully mapping the untranscribed audio to a sequence of CLIP subword tokens would felicitate unsupervised speech recognition. 

In the end, we move beyond CLIP, trained with parallel image-text data as the feature extractor. We extend our segmental SpeechCLIP framework to enable the utilization of image and text encoders that are trained with only unimodal data. Unimodal data, e.g., text or image, are much more abundant than parallel image-text data. For example, images with parallel text captions are a small subset of online text and image data. This allows us to leverage models that are trained on a much more significant amount of data than CLIP. 
We also explore techniques to utilize text encoders to improve speech systems without any parallel image data. 
We analyze how much of the semantic knowledge is learned due to the parallel images or multiple parallel spoken descriptions. 

We use the SpokenCOCO dataset~\cite{hsu2021text} for training and evaluating the proposed method. On the image-speech and speech-image retrieval task, our model significantly outperforms the cascaded variant of SpeechCLIP. In the end, we show competitive performance on the Zerospeech 2021 semantic similarity task~\cite{nguyen2020zero}.  

The main contributions of this work are as follows:

\begin{itemize}
    \item We show our segmental framework enables better utilization of pretrained text encoder 
    \item Modifications to contrastive loss to reduce the computational requirement of training audio-visual systems
    \item Our framework can utilize text and image encoders trained with unimodal data only
    \item Extensive analysis of different hyperparameters of system
    \item We show it is possible to distill semantic information from pretrained text encoders to speech encoders 
    \item We can also leverage pretrained text encoders without parallel image-speech data
\end{itemize}

The rest of the paper is organized as follows: First, we present the related work in Section~\ref{sec:rel}. Then, we present our proposed segmental speechCLIP in Section~\ref{sec:SegSpeechCLIP}. Then, the experimental setup and results are detailed in Section~\ref{sec:exp}. We then discuss how to utilize unimodal pretrained models to improve the performance of visually grounded speech models in Section~\ref{sec:unim}. We analyze the semantic information present in segmental speechCLIP in Section~\ref{sec:semantic}. We explore the possibility of building audio-only systems in Section~\ref{sec:audioonly}. We conclude the chapter in Section~\ref{sec:conc3}.

\section{Related work} \label{sec:rel}

Visually grounded speech systems have become very popular in recent years. Joint image and text learning systems popularized grounding images with language i.e. text. A lot of the VGS systems are based on earlier works on images and text. Another emerging trend is to utilize a pretrained image and/or text encoder and distill knowledge into the speech encoder. Chrupala~\cite{chrupala2022visually} provides an extensive history of visually grounded models including datasets, architectures, downstream tasks, and evaluation techniques. 

Synnaeve et al.~\cite{synnaeve2014learning} proposed one of the earliest joint visually grounded models which mapped image and speech to a common space with cosine-distance loss. Harwath et al.~\cite{harwath2016unsupervised} scaled up the joint learning approach with larger models and larger datasets. They also modified the training objective and introduced a contrastive margin loss. The contrastive loss aims to increase the similarity between matching image-speech pairs compared to the random pairs of image-speech at least by a prespecified margin.

Earlier models relied on convolutional neural networks~\cite{harwath2016unsupervised,harwath2017learning} as the speech encoders. Recent works explore different architectures. The attention mechanism in particular has become a popular choice. Chrupała et al.~\cite{chrupala2017representations} used a gated RNN with attention for learning speech representation. Harwath et al. explored residual networks as the speech encoders~\cite{harwath2018jointly,harwath2017learning} and further explored the use of vector quantized residual networks~\cite{harwath2017learning} for learning discrete speech units. Peng et al. proposed transformer-based models~\cite{peng2022fast,peng2022fastvgsplus}. 

While the initial models focused on learning both the image and speech encoders from scratch, now model distillation has become standard practice where a large pretrained image encoder is used to guide the audio model~\cite{gelderloos2016phonemes,harwath2017learning,alishahi2017encoding,harwath2019learning}. Harwath et al.~\cite{harwath2018jointly} explored the impact of pretraining the image model on the system performance. They discovered that pretraining the image model improves the model performance on the downstream retrieval task. While supervised pretraining is the most helpful, unsupervised pretraining was not far below supervised training. Pretraining allows the model to look at a wide variety of data and thus helps improve performance.

Recently, there have been methods that utilize large self-supervised models trained on large amounts of unlabeled audio data as feature extractors. Peng et al.~\cite{peng2022fast,peng2022fastvgsplus} utilized the Wav2Vec2 model as the feature extractor. ~\cite{speechclip2022} used HuBERT as the feature extractor. Peng et al.~\cite{peng2022fastvgsplus} extended the model to use the MLM loss to utilize unlabeled audio data. Recent models such as SpeechCLIP~\cite{speechclip2022} had most of their parameters in the frozen feature extractors such as CLIP and HuBERT and only a small percentage (around 1-2\%) of the parameters were learned during the training.

However, the majority of these models focus on leveraging the pretrained image and speech encoders, and the pretrained text encoders are not utilized for training VGS systems. Guzhov et al.~\cite{guzhov2022audioclip} proposed AudioCLIP to leverage both image and text encoders for the Environmental Sound Classification task. However, the pretraining setup requires manual labels for the audio encoder. Zhao et al.~\cite{zhao2021connecting} focused on aligning audio and text for audio classification tasks by using the vision as the pivot. They leveraged both the image and text encoders.  
Cascaded SpeechCLIP~\cite{speechclip2022} used an attention-based module to downsample and learn K word-like units from the audio. Their main goal was to learn localized word information and distill semantic information into the speech encoder. Segmental SpeechCLIP~\cite{bhati23_interspeech} improved the unsupervised word process by using a segmental speech encoder. Segmental SpeechCLIP doubles the retrieval performance and achieves good performance of the semantic similarity task. 


\section{Segmental SpeechCLIP} \label{sec:SegSpeechCLIP}
The CLIP model is trained with a large amount of paired image-text data. CLIP uses two encoders for processing images and text separately and learns to align semantically similar images and text captions. The features extracted from CLIP transfer well to other computer vision tasks. We aim to utilize both the text and image encoder to learn speech representations. By cascading the output of the segmental speech encoder with the CLIP text encoder, we aim to induce semantic information in the speech encoder.

\begin{figure}
    \centering
    \includegraphics[width=0.95\columnwidth]{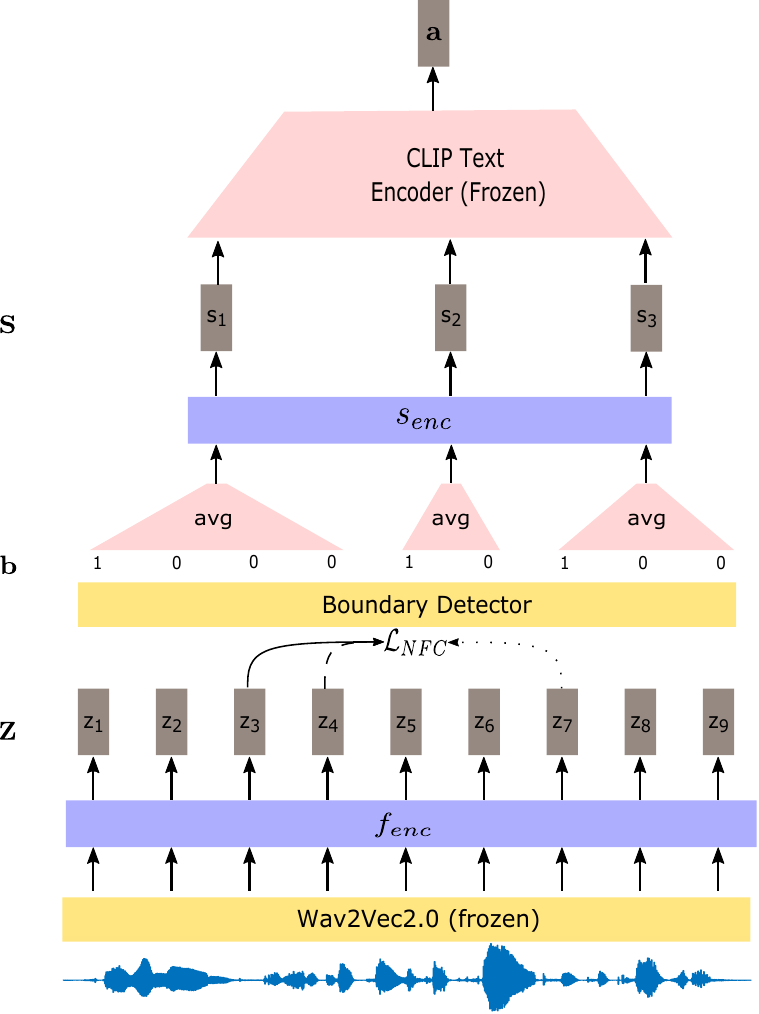}
    \caption{Overview of the Segmental speech encoder. }
    \label{fig:SSE}
    \vspace{-5mm}
\end{figure}

The main difference between our proposed method and previously proposed approaches is the word extraction process from the utterances. The segmental speech encoder used for word extraction is summarized in Figure~\ref{fig:SSE}. For the audio encoder, we first use frozen Wav2vec2 to extract audio frame-level features. A trainable segmental audio encoder then extracts sub-words from the frame-level features. The frozen CLIP text encoder generates sentence embeddings from the sub-words. A frozen CLIP image encoder is used for extracting image embeddings. We describe the various components of the segmental speech encoder in detail below. 

\subsection{Next frame classifier}

Let the sequence $\mathbf{X} = (x_1,x_2,...,x_T)$ represent a waveform. We use a frozen Wav2vec2 followed by a feed-forward network to extract frame level features $\mathbf{Z}(\in \mathbb{R}^{p\times L}) = (\zvi{1},\zvi{2},...,\zvi{L})$ at low frequency. Each p-dimensional vector $\zvi{i}$ corresponds to a 25 ms audio frame extracted with a 20 ms shift. Given frame $\zvi{t}$, the encoder ($f_{enc}$ in Fig. \ref{fig:SSE}) tries to classify the next frame $\zvi{t+1}$ correctly within a set of $K+1$  representations $\Tilde{\zvi{}} \in \mathcal{Z}_{t}$, which include $\zvi{t+1}$ and $K$ negative examples, randomly sampled from the same utterance, as

\begin{equation}
    \mathcal{L}_{\mathrm{NFC}} = -\log \frac{\exp(\mathrm{sim}(\zvi{t},\zvi{t+1}))}{\sum_{\Tilde{\zvi{}} \in \mathcal{Z}_{t} } \exp(\mathrm{sim}(\zvi{t},\Tilde{\zvi{}} ))}
\end{equation}
where $\mathrm{sim}(\mathbf{x},\mathbf{y}) =  \frac{\mathbf{x}\mathbf{y}^{T}}{\Vert \mathbf{x} \Vert \Vert \mathbf{y} \Vert} $ denotes the cosine similarity.

\subsection{Boundary detector}
We use the boundary detector from~\cite{bhati21_interspeech}, which compares the adjacent frames and output a boundary if the similarity between the adjacent frame falls below a threshold. The boundary detector outputs a sequence of ones and zeros, each one indicating if there is a boundary change at that timestep. We generate the segment representations by feeding the average of constituting frames in the segment through a segment encoder, $\mathrm{s_{\mathrm{enc}}}$. We use the vectorized computation method from ~\cite{bhati21_interspeech} for a faster segment representation calculation. 
After the boundary detection stage the feature sequence $\mathbf{Z} = (\zvi{1},\zvi{2},...,\zvi{L})$ is segmented into disjoint contiguous segments $\mathbf{S} = (\svi{1},\svi{2},...,\svi{M})$

\subsection{Directly learning the CLIP sub-word representations}
In the cascaded SpeechCLIP approach, the audio encoder generates keyword embeddings. Then, argmax is used to find the index of subword embedding in the CLIP vocabulary closest to the keyword. The corresponding subword embedding is used as the keyword embedding. This process uses the argmax operator, which is non-differentiable. Straight-through gradient estimator is used for training the model. 

Here, we directly generate subword embeddings via the segmental audio encoder. These embeddings are fed into the text encoder bypassing the pretrained vocabulary in the CLIP model. This way, our model can be trained without straight-through estimators. However, this may not generate exact embeddings from the vocabulary. We pass the segment embeddings $\mathbf{S} = (\svi{1},\svi{2},...,\svi{M})$ through the CLIP text encoder to generate the audio embedding, $\mathbf{a}$.
\subsection{Retrieval loss}
Typically, both audio and image encoders are trained in visually grounded models. The image embedding should be closest to the corresponding audio embedding, and the audio should be closest to the corresponding image embedding from a pool of negative examples. The loss is the sum of the two losses.
Since we are not training the image encoder, we train the audio encoder to pick the image embedding of the paired image from a set that contains negative examples. It can be considered a classification problem where we classify the paired image embedding from a set with negative examples given the audio embedding. More information on the negative sampling process can be found in the next section. The retrieval loss is given as: 

\begin{equation}
   \mathcal{L}_{\mathrm{RET}} =  -log \frac{\exp (\mathbf{a_k}\mathbf{i_k}^{T} / \tau)}{ \sum_{i' \in \mathcal{I}_{k} } \exp (\mathbf{a_k}\mathbf{i^{'T}} / \tau) }
\end{equation}
where $\mathbf{a_k}$ is the output of the CLIP text encoder, $\mathbf{i_k}$ is the output of the CLIP image encoder and $\tau$ is the temperature. 

Our model has multiple components, and we train our model progressively. We begin by training the frame-level encoder for a few steps and then add the retrieval loss. The two losses are trained together for the first epoch. For the rest of the training, only the retrieval loss is optimized. The overall loss of the model is given as
\vspace{-2mm}
\begin{equation}
     \mathcal{L}= 
\begin{cases}
    \mathcal{L}_{\mathrm{NFC}},& \text{if } \text{step}\leq 100 \; \& \; \text{epoch} = 1\\
    \mathcal{L}_{\mathrm{NFC}} + \mathcal{L}_{\mathrm{RET}},              & \text{if } \text{step}\geq 100 \; \& \; \text{epoch} = 1 \\
    \mathcal{L}_{\mathrm{RET}},              & \text{otherwise} 
\end{cases}
\end{equation}

\subsection{Negative sampling}
Negative sampling is an important part of contrastive loss. It helps us prevent model collapse. We make the following two changes to the negative sampling process. 

\subsubsection{Disentangling batch size and negative examples}
Typically, the negative samples are sampled from the batch. This, unfortunately, ties the number of negative examples with the batch size. To increase the number of negative examples, we must increase the batch size, which is not always possible on smaller GPUs. 

Since we are using frozen pretrained CLIP, i.e., image representation stays the same throughout training. We load all the image representations in advance and sample negative examples from them. This way, we can increase the number of negative examples without increasing the batch size. This is impossible when training the image encoder, as the image representation changes after every update. 

\subsubsection{Hard negative mining through clustering}

The quality of the negative samples impacts the contrastive loss, and sampling better negative examples has been an active research direction. We want to find the closest examples for each entry in the batch and then contrast them. Since the image embeddings are fixed, we can find the closest examples before training the audio encoder. Here we use a simple clustering-based technique to sample harder negative examples. 

We first cluster the image embeddings for the dataset into K subsets and store the cluster index for each image embedding. During training, we find the cluster index for each embedding and sample up to 512 examples from that cluster. These form the hard negative examples. The rest of the examples are randomly sampled. Again this is possible because the image embeddings do not change during training. 

\section{Experiments} \label{sec:exp}

\begin{table*}
    \centering
    \caption{Recall scores for image-speech retrieval on SpokenCOCO 5k test set}
    \label{tab:ret_perf}
    {
    \begin{tabular}{cccccccccc}
            \toprule
            & & Image & & & Speech & & & Mean &\\ \cmidrule{2-10}
         & R@1 & R@5 & R@10 & R@1 & R@5 & R@10 & R@1 & R@5 & R@10\\ \cmidrule{2-10}
        Parallel Speech CLIP & 35.8 & 66.5 & 78.0 & 50.6 & 80.9 & 89.1 & 43.2 & 73.7 & 83.6\\
        Cascaded SpeechCLIP & 6.4 & 20.7 & 31.0 & 9.6 & 27.7 & 39.7 & 8 & 24.2 & 35.6 \\
        Seg. SpeechCLIP & 28.2 & 55.3 & 67.5 & 28.5 & 56.1 & 68.9 & 28.4 & 55.7 & 68.2\\ \bottomrule
    \end{tabular}
    }
    \vspace{-4mm}
\end{table*}

\subsection{Dataset}
We use the following datasets to validate our models: 

\begin{itemize}
    \item \textbf{SpokenCOCO}: SpokenCOCO is a spoken version of the MSCOCO captioning dataset~\cite{hsu2021text}. Spoken captions are collected by displaying the text captions to a person and having them read them aloud. Each image in the dataset is paired with five spoken captions. SpokenCOCO contains 123k images and 742 hours of speech from 2353 speakers. Utterances, on average, are around 4 seconds long and contain 10.45 words. We follow the SpeechCLIP~\cite{speechclip2022} for train/test splits. We use a much smaller split of validation to save time during training. 
    
    \item \textbf{sSIMI}: sSIMI is the spoken version of 13 existing text-based semantic similarity and relatedness tests~\cite{dunbar2021zero}. These datasets contain word pairs and semantic similarity scores. All scores were normalized on a 0-10 scale. All the pairs that had a word not in the Librispeech were discarded. One of the datasets, mturk-771, was used for creating the dev set, and the rest 12 were used for making the test set. There is no overlap in the dev and test set. Each set contains two subsets of audio files: one synthetic and one natural. To create the synthetic subset: a speech synthesizer is used. For the natural subset: audio segments are extracted from Librispeech utterances. The natural subset is smaller than the synthetic subset as the pairs containing words not in Librispeech dev and test sets were discarded. The synthesized subset contains 9744 and 705-word pairs for the test and dev sets, respectively, and the LibriSpeech subset contains 3753 and 309 pairs for the test and dev sets.
\end{itemize}

\subsection{Model architecture} 
In our experiments, the wav2vec2~\cite{baevski2020wav2vec} model and the CLIP model are frozen. We use them as feature extractors. The next frame classifier is a three-layer feed-forward network with 1024 hidden units. The segment encoder contains two convolutional layers with 1024 filters with kernel size three, followed by a feed-forward network with two layers with either 512/768 hidden units for small/large CLIP models. The segmental speech encoder contains approximately 10 million parameters. We use Adam optimizer with a 2e-5 learning rate and a batch size of 21. We decay the learning rate by a factor of 0.95 every three epochs. All the experiments are conducted on a single 11GB GPU.   

\subsection{Retrieval performance}
In this section, we evaluate the segmental SpeechCLIP on the image-speech retrieval task to measure how well we can align speech and image embeddings. As shown in Table~\ref{tab:ret_perf} our proposed model significantly outperforms the cascaded SpeechCLIP model.
We almost doubled the performance of cascaded SpeechCLIP. Segmental SpeechCLIP has a slightly lower number of trainable parameters. We believe this is due to the improved word-discovery process in segmental SpeechCLIP.

We use Wav2vec2.0 as the feature extractor for speech, whereas cascaded SpeechCLIP uses Hubert~\cite{hsu2021hubert} as the feature extractor. 
Another big difference is the feature extraction process; we use the features from a single layer, i.e., layer 11 in the Wav2vec2 transformer encoder, where cascaded SpeechCLIP learns weights to combine the transformer encoder's hidden representations. A weighted combination of layer-wise features from wav2vec2/Hubert tends to work better than single-layer features on downstream tasks~\cite {yang21c_interspeech,tsai2022superb}. We only use features from a single layer due to GPU memory constraints. 

Our approach shows the potential for utilizing pretrained text encoders for improving VGS systems. However, the performance is still lower than the parallel variant of SpeechCLIP, which extracts global representations from speech. We hypothesize that the segmentation process and the passing of the segmented speech through the CLIP text encoder lose information.  

\subsection{Impact of hard mining through clustering}

We proposed to use clustering for mining hard negative examples. We explore if this change helps the learning process. We train a system where all the negative examples are sampled randomly and the other where clustering is used for selecting the hard negative examples. Both cases use the same number of negative examples. As seen from ~\ref{tab:Clust_samp}, hard mining via clustering helps the learning process. 

\begin{table}[h!]
    \centering
    \caption{Average retrieval on SpokenCOCO test set. ``with" and ``without" indicate use/lack of clustering for hard mining negative examples.}
    \begin{tabular}{cccc} 
        \toprule
         & R@1 & R@5 & R@10  \\ \cmidrule{2-4}
        with & 26.1 & 52.2 & 64.8 \\
        without & 22.4 & 48.6 & 61.1 \\ \bottomrule
    \end{tabular}
    \label{tab:Clust_samp}
\end{table}

\subsection{Impact of initial word boundaries}
Unsupervised word segmentation has been a growing research area. Recent state-of-the-art solutions utilize multimodal (paired speech, image) data~\cite{peng2022word}. Our segmental encoder segments the audio data in sub-word-like segments. We experiment with whether using the word boundaries from an existing word segmentation system can be useful. We use the VG-Hubert model for extracting the initial word boundaries~\cite{peng2022word}. VG-Hubert achieves the best word segmentation performance on TIMIT and Buckeye datasets. We insert the VG-Hubert boundaries in between the boundaries generated by the segmental speech encoder.

As seen from Table~\ref{tab:init_bound}, using word boundaries has no or a little negative impact on retrieval performance. Either the word boundaries do not help the retrieval task, or our insertion process is not optimal. We plan to explore more ways of utilizing the VG-Hubert boundaries in the segmental SpeechCLIP model. 
\begin{table}[h!]
    \centering
    \caption{Average retrieval on SpokenCOCO test set. ``with" and ``without" indicate use/lack of initial word boundaries.}
    \begin{tabular}{cccc}
        \toprule
         & R@1 & R@5 & R@10  \\ \cmidrule{2-4}
        with & 22.4 & 48.6 & 61.1 \\
        without & 23.1 & 49.2 & 61.5 \\ \bottomrule
    \end{tabular}
    \label{tab:init_bound}
    \vspace{-4mm}
\end{table}
\subsection{Impact of model size}
The CLIP model is used as a feature extractor for images and for extracting the final audio representations. We analyze the Impact of CLIP model size on retrieval performance. We use the CLIP small model (ViT-B/32) with 250 million parameters and the large CLIP model (ViT-L/14) with 422 million parameters. 
The segmental speech encoders used in the two cases are very similar. For the large CLIP model, the SSE generates 768-dimensional representations; for the small CLIP model, the output dimension is 512. 

As evident from Table~\ref{tab:CLIP_size}, the large model helps the retrieval performance. The observation is similar to ~\cite{speechclip2022}, where the system with large CLIP models outperformed the one with smaller CLIP models. 
\begin{table}[h!]
    \centering
    \caption{Average retrieval on SpokenCOCO test set. Small/Large denotes the CLIP model size.}
    \begin{tabular}{cccc}
        \toprule
         & R@1 & R@5 & R@10  \\ \cmidrule{2-4}
        Small & 26.1 & 52.2 & 64.8 \\
        Large & 28.3 & 55.7 & 68.2 \\ \bottomrule
    \end{tabular}
    \label{tab:CLIP_size}
    \vspace{-4mm}
\end{table}
\subsection{Enforcing proximity to CLIP vocabulary embeddings}
In our approach, we bypass the vocabulary of the CLIP text encoder and directly learn the input representations. However, we may learn representations not belonging to the CLIP vocabulary. We enforce that the generated embeddings should be close, ideally the same, as those in the CLIP vocabulary. If successful, it will allow us to transcribe audio as a sequence of CLIP subword tokens thus effectively achieving unsupervised ASR. We experiment with two approaches:

\subsubsection{Via regularization}

In this approach, we maximize the similarity between the segment embeddings $\mathbf{S} = (\svi{1},\svi{2},...,\svi{M})$ and CLIP vocabulary embeddings, V. We first compute the cosine similarity between the $j^{th}$ normalized word embedding and the $k^{th}$ subword embedding ($e_{v}$) from the text encoder, e.g., CLIP as
\begin{equation}
    c_{jk} = \mathrm{cos}(\svi{j},e_k)
\end{equation}

For an utterance with $M$ segments, the regularizing loss is computed as 

\begin{equation}
    \mathcal{L}_{reg} = -\frac{1}{M}\sum_{j=1}^{M} \log{(\max_{k} c_{jk} )}
\end{equation}

The final loss of the model is given as 
\begin{equation}
    \mathcal{L}_{F} = \mathcal{L} + \lambda \mathcal{L}_{reg}
\end{equation}
Where $\lambda$ is the regularization weight.

\begin{table}[h!]
    \centering
    \caption{Average retrieval vs the regularization weight ($\lambda$) on SpokenCOCO test set.}
    \begin{tabular}{cccc}
        \toprule
         & R@1 & R@5 & R@10  \\ \cmidrule{2-4}
        0.0 & 26.1 & 52.2 & 64.8 \\
        0.1 & 18.1 & 40.7 & 53.8 \\
        0.5 & 14.0 & 34.6 & 47.5 \\ 
        1.0 & 12.2 & 32.5 & 45.5 \\ \bottomrule
    \end{tabular}
    \label{tab:Retvsreg}
\end{table}

Table~\ref{tab:Retvsreg} shows the effect of regularization weight on the retrieval performance. As evident from the Table, the retrieval performance decreases with an increase in the regularization weight. Please note that even the regularized systems outperform the cascaded SpeechCLIP. 

\subsubsection{Vector Quantization}

Here, we map the segment embedding to CLIP's embeddings via vector quantization. We can choose the subword embedding with the highest similarity from the text encoder vocabulary as

\begin{equation}
    h_{k}^{\mathrm{hard}} = e_{v^{*}}, \text{where} \; v^{*} = \underset{1\leq v \leq V}{\mathrm{argmax}}  \; c_{jk}
\end{equation}

However, the argmax operator is not differentiable, so we use a gradient straight-through estimator to train the model. 

\begin{equation}
    h_{k} = h_k^{\mathrm{soft}} + \mathrm{sg}(h_{k}^{\mathrm{hard}} - h_k^{\mathrm{soft}})
\end{equation}

Where sg denotes the stop gradient operator. In the forward pass $h_{k}^{\mathrm{hard}}$ is used and in the backward pass $h_{k}^{\mathrm{hard}}$ is used. $h_{k}^{\mathrm{soft}}$ can be computed as follow

\begin{equation}
    h_{k}^{\mathrm{soft}} = \left[ e_{1}...e_{V}\right] \mathrm{softmax} (\left[ c_{k1}...c_{kV} \right]^{T} / \tau)
\end{equation}

Where $\tau$ is the temperature. 

Table~\ref{tab:retvsVQ} shows the impact of using vector quantization on retrieval performance. With vector quantization, the performance of the system goes down. The VQ segmental speechCLIP still outperforms the Cascaded speechCLIP significantly. 

\begin{table}[h!]
    \centering
    \caption{Impact of quantization on the average retrieval on SpokenCOCO test set.}
    \begin{tabular}{cccc}
        \toprule
         & R@1 & R@5 & R@10  \\ \cmidrule{2-4}
        No VQ & 26.1 & 52.2 & 64.8 \\
        VQ & 11.8 & 38.6 & 56.6 \\ \bottomrule
    \end{tabular}
    \label{tab:retvsVQ}
\end{table}

The retrieval performance is lower than the unconstrained system performance in both the regularization and quantization cases.  
We think the segment generation process is not optimal, and forcing the generated segment embeddings to match CLIP embeddings further increases the optimization process's difficulty, resulting in lower retrieval performance. 


\section{Utilizing unimodal pretrained models} \label{sec:unim}

In our experiments, we utilize CLIP as the image and text encoder. CLIP is trained on parallel image and text data, which is less abundant than just text or image data. In this section, we explore if we can utilize SSL models trained with only text or image data to improve system performance.

\begin{table*}[h!]
    \centering
    \caption{Recall scores for image-speech retrieval on SpokenCOCO 5k test set}
    \label{tab:LLMtarget}
    {
    \begin{tabular}{ccccccccccc}
            \toprule
            Image encoder & Text encoder & &Image & & & Speech & & & Mean \\ \cmidrule{1-11}
         & & R@1 & R@5 & R@10 & R@1 & R@5 & R@10 & R@1 & R@5 & R@10\\ \cmidrule{1-11}
        CLIP small & CLIP small & 25.4 & 51.6 & 63.9 & 26.7 & 52.8 & 65.6 & 26.1 & 52.2 & 64.8\\
        CLIP small & mpnet & 29.5 & 57.3 & 68.9 & 30.3 & 57.3 & 69.6 & 29.9 & 57.3 & 69.3 \\
        CLIP large & CLIP large & 28.2 & 55.3 & 67.5 & 28.5 & 56.1 & 68.9 & 28.4 & 55.7 & 68.2\\
        CLIP large & mpnet & 32.7 & 61.3 & 72.9 & 33.5 & 60.8 & 72.3 & 33.1 & 61.1 & 72.6 \\
        DINO & mpnet & 19.7 & 44.7 & 57.6 & 20.2 & 44.4 & 57.5 & 20.0 & 44.5 & 57.5 \\
    \bottomrule
    \end{tabular}
    }
\end{table*}

\subsection{Pretrained LLM as additional loss}

One of the main motivations for using the CLIP text encoder is to distill semantic information in the speech encoder. We explore if adding an additional language model pretrained on a large amount of text data can bring in further improvements. 

\begin{figure}[h!]
    \centering
    \includegraphics[width=0.95\columnwidth]{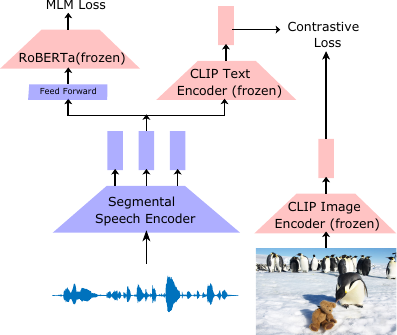}
    \caption{Combining the Segmental Speech encoder with pretrained large language model}
    \label{fig:mlm_overview}
\end{figure}

We add an extra pretrained language model, RoBERTa, to our model. RoBERTa is trained with masked language modeling objectives on large amounts of text data. The final architecture is shown in Figure~\ref{fig:mlm_overview}. We mask some segment embeddings, and the MLM tries to predict the masked embeddings. There might be a difference in the embedding space of the pretrained MLM and CLIP text encoder. The segment embeddings, $\mathbf{S}$, are passed through a small feed-forward network before the pretrained MLM. The feed-forward network tries to project the segments into the embedding space of the pretrained MLM. Again we directly use the embeddings as input to the model instead of quantizing. This also opens up the possibility of utilizing speech-only data for training the SSE. Table \ref{tab:MLM_loss} shows the retrieval performance with and without using the RoBERTa model. We also experiment with adding a larger version of the RoBERTa model. The additional MLM has little to no improvements. There could be several factors for this: i) we need to optimize the MLM loss parameters such as the masking probability or the mask span, ii)  maybe there is too much mismatch between the  CLIP embedding space to RoBERTa embedding space and a simple feed-forward network is not enough. 

\begin{table}[h!]
    \centering
    \caption{Average retrieval on SpokenCOCO test set. Small/Large denotes the CLIP/RoBERTa model size.}
    \begin{tabular}{ccccc}
        \toprule
        CLIP size & RoBERTa size & R@1 & R@5 & R@10  \\ \midrule
        Small & None & 26.1 & 52.2 & 64.8 \\
        Small & Small & 26.0 & 52.2 & 65.1 \\ \midrule
        Large & None & 28.3 & 55.7 & 68.2 \\ 
        Large & Large & 29.4 & 56.2 & 68.6 \\ \bottomrule
    \end{tabular}
    \label{tab:MLM_loss}
    \vspace{-4mm}
\end{table}

\subsection{Pretrained LLM as targets}

We use pretrained CLIP to distill knowledge into the speech encoder. Pretraining CLIP still requires paired image-text data. We want to explore if we can also distill knowledge from text and image encoder pretrained with unimodal data. We experiment with various image and text encoder combinations for the retrieval task. 
For the text encoder, we use all-mpnet-base-v2~\cite{mpnetv2}, a mpnet~\cite{song2020mpnet} based sentence transformer model from Huggingface~\cite{wolf-etal-2020-transformers} and the CLIP text encoder as the possible options. We experiment with DINO~\cite{caron2021emerging} and CLIP image encoder for the image encoder.

The all-mpnet-base-v2 model uses the pretrained mpnet model~\cite{song2020mpnet} and finetunes it on the 1 Billion sentence pairs dataset. The model is trained to predict the pairs of sentences among randomly sampled negative sentences. DINO~\cite{caron2021emerging} consists of teacher and student models. The student and teacher models are presented with different views of the same image. The student is trained to predict the output of the teacher model. The teacher model parameters are updated with the exponential moving average of the student weights. 

Table~\ref{tab:LLMtarget} shows the performance of various SSL image and text encoders. We experiment with both small and large CLIP models, and using the mpnet model consistently improves the retrieval performance. The mpnet models are trained on more text data than the CLIP model. Using DINO as the image encoder decreases the performance. The DINO is trained on a much smaller amount of data. The size of embeddings for DINO is also smaller than the CLIP models. DINO uses 384-dimensional embeddings, whereas CLIP uses 512/768-dimensional embeddings. It would be interesting to see when the unimodal models, such as DINO, and mpnet, are trained on the same amounts of data as the CLIP model. It will allow us to examine whether unimodal systems can truly replace multimodal for knowledge distillation. We cannot train any of these models due to computational constraints and only use them as feature extractors.

\begin{table*}[h!]
    \centering
    \caption{Semantic similarity scores on the Zerospeech 2021 sSIMI task}
    \begin{tabular}{cccccc} \toprule
         &Budget &\multicolumn{2}{c}{dev} & \multicolumn{2}{c}{test}  \\ \cmidrule{2-6}
         & & synthetic & Librispeech & synthetic & Librispeech \\ \cmidrule{2-6}
        VG baseline & 72 & 9.65 & 12.61 & 9.71 & 0.16 \\ 
        VG baseline & 160 & 9.60 & 15.09 & 9.99 & -0.10 \\ \midrule

        FaST-VGS+~\cite{peng2022fastvgsplus} & 468 &23.07 & \textbf{23.10} & 15.10 & 14.32 \\
        Seg. SpeechCLIP & 72 & \textbf{28.79} & 16.80 & \textbf{19.60} & \textbf{15.69} \\  
        \midrule

        Phone topline & 1536 & 9.86 & 16.11 & 12.23 & 20.16 \\
        \bottomrule
    \end{tabular}
    \label{tab:sSIMI}
\end{table*}

\section{Semantic representation learning} \label{sec:semantic}

One of the motivations for utilizing the text encoder in the pretrained CLIP was to learn semantic representations. We use the Zerospeech 2021 challenge semantic similarity task, sSIMI, to evaluate the representations' quality. This task aims to compute the similarity between representations of pairs of words and compare it with similarity scores assigned by human annotators. 
We use the sSIMI dataset for evaluating the system performance.
More details about the task and the evaluation can be found in the challenge paper~\cite{dunbar2021zero}.

As seen in Table~\ref{tab:sSIMI}, we perform competitively with a state-of-the-art method. However, there are a few key differences in the methodologies. Our model has fewer parameters and less training time than FaST-VGS+. 
FaST-VGS+ relies on pretrained R-CNN to generate bounding boxes for the objects in the image; our model does not.
FaST-VGS+ can leverage speech-only data via a Masked language modeling task which needs to be improved in our approach. FaST-VGS+ is trained on SpokenCoCo and Librispeech, whereas we only use SpokenCOCO. This might explain the lower performance on Librispeech test data.   

\section{Without image encoder} \label{sec:audioonly}

Audio-visual systems allow us to ground audio representations with visual representations. These systems learn from the co-occurrence of concepts between the two modalities. SpokenCOCO contains multiple spoken descriptions of the same image. All these utterances describe the same underlying concept, even if they use different words. We want to explore whether it is possible to utilize pretrained text models and learn semantic information without using the image information. 

\begin{figure}[h!]
    \centering
    \includegraphics[width=0.95\columnwidth]{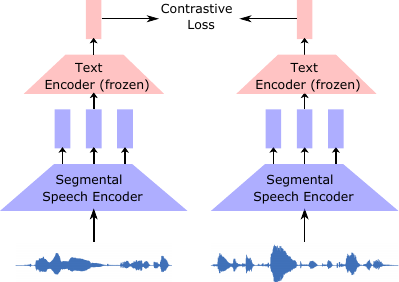}
    \caption{Distillation without image encoder. The two speech inputs are semantically related, e.g., different descriptions of the same image.}
    \label{fig:audio_only}
\end{figure}

We use a similar architecture to the audio-visual systems. Figure \ref{fig:audio_only} shows the overview of the proposed approach. The model contains two branches; we feed two spoken descriptions of the same image to the left and right branches. We do not use the image and only use the audio. 

The Segmental Speech encoder shares parameters across the branches. We experiment with using mpnet as the text encoder on both sides, mpnet on one side, and CLIP text encoder on one side. Since mpnet and CLIP text encoder expect different embedding dimensions, we add a small neural network in of the branches, e.g., right in the figure to map the output of the SSE in the required dimension.
The model is trained with contrastive loss to maximize the similarity between semantically similar audio pairs over random audio pairs.

\subsection{Semantic audio retrieval task}

We evaluate this system on the retrieval task similar to before. Now instead of retrieving images from audio and vice-versa, the objective is to retrieve the semantically similar spoken description. We use the val/test set from the SpokenCOCO dataset for generating the semantic pairs for evaluation. The same utterance pairs are used across all evaluations. 
We compare models trained with image-audio and audio-only models on the semantic audio retrieval task. The number of trainable parameters is almost the same in the two settings. We use the SpokenCOCO dataset for training the models.

The image-audio models are trained with a large number of negative examples, i.e., 1024. We want to match the number of negative examples for audio-only models. Since the audio encoders evolve during training, we can not use the disentanglement trick to increase negative examples. Now the number of negative examples is tied to the batch size, and we can not increase the batch size due to computational limitations. We try increasing the number of negative examples by storing previous batches. We keep the sentence embeddings of earlier batches and use them as negative examples. However, this does not work, and the model fails to learn meaningful representations. We hypothesize our batch size is too small, so we need to keep outputs from a distant past as the negative examples, but the model has changed considerably by then. For example, our batch size is 15, and the speech-image system was trained with 1024 negative examples, so to generate 1024 examples, we need to store 68 previous batches. The model weights update after each batch, and after 68 batches, the stored negative examples might not be useful. We train new image-audio systems with batch size and the number of negative examples as 15.

Since both sides in the audio-only model can accept audio input and generate features. We explore which branch is better for generating features for downstream tasks. We experiment with features from left, and right and the concatenation of the two branches. We evaluate the feature on the semantic audio retrieval task. As shown in Table~\ref{tab:Audio_only_lr}, the concatenation of the features works almost as well as the features from the right branch, but they are double in dimensionality. For future evaluation, we use the features from the right branch. 

\begin{table}[h!]
    \centering
    \caption{Impact of features extracted from left or the right branch on the semantic audio retrieval task on SpokenCOCO validation set.}
    \begin{tabular}{cccc}
        \toprule
         & R@1 & R@5 & R@10  \\ \cmidrule{2-4}
        Right & 16.8 & 38.8 & 50.5 \\
        Left & 14.9 & 35.5 & 47.6 \\ 
        Concat (Right,left) & 16.9 & 38.9 & 50.6 \\ \bottomrule
    \end{tabular}
    \label{tab:Audio_only_lr}
\end{table}

Next, we compare the audio-image and audio-audio models. As seen in Table~\ref{tab:aavsai}, the audio-audio models perform worse on the semantic audio retrieval task. This shows that grounding with vision helps distill more semantic information into the audio encoder. It also shows that not all the semantic information comes from image-audio pairs; some comes from pairs of spoken audio captions. It is very promising, as it opens up the possibility of distilling semantic information into audio encoders without any parallel image data. 
The audio-only model using mpnet text encoder on both sides outperforms the mpnet-CLIP model. These observations are consistent with previous observations in this paper. 

\begin{table}[h!]
    \centering
    \caption{Comparison of the audio-audio vs. audio-image systems on the semantic audio retrieval task on SpokenCOCO validation set.}
    \begin{tabular}{cccccc}
        \toprule
         & Image enc. & Text enc. & R@1 & R@5 & R@10  \\ \cmidrule{2-6}
        Audio-image & CLIP & mpnet & 21.4 & 43.5 & 55.4 \\
        Audio-audio & None & mpnet, CLIP & 14.2 & 34.2 & 46.4 \\ 
        Audio-audio & None & mpnet, mpnet & 17.1 & 38.8 & 50.5 \\
        \bottomrule
    \end{tabular}
    \label{tab:aavsai}
    \vspace{-4mm}
\end{table}

\section{Conclusions and future work} \label{sec:conc3}
Our segmental SpeechCLIP framework allows us to leverage pretrained text encoders for improving the VGS systems. Our method doubles the performance of recent attempts that utilize the pretrained text encoders. It shows the importance of building better word discovery systems, as one of the major differences between the two approaches was the word discovery process. We modify the contrastive loss to disentangle batch size and the number of negative examples. This allows us to use a larger number of negative examples without increasing the computational requirements. We also proposed a clustering-based hard mining technique to improve the quality of the negative examples. The improvement in the quality of the negative improves the downstream retrieval performance. 

Our experiments show that using a bigger text encoder improves performance. Although, increasing scale by using multiple text encoders in parallel does not increase the system's performance. We then extend our framework to utilize pretrained text encoders when only speech data is available. We show that on the semantic audio retrieval task, just using the multiple spoken descriptions of the same image works almost as well as using images with spoken descriptions. 
Our experiments show that a pretrained text encoder allows us to distill semantic information into the speech encoder. Our methods outperform the existing ones on the Zerospeech 2021 semantic similarity task.  

In the future, we would like to explore the application of unsupervised ASR by mapping the unlabeled audio to a sequence of CLIP vocabulary tokens.
Currently, we used features from a fixed layer from a pretrained feature extractor, i.e., wav2vec2, as the input for our model. 
We want to explore a weighted combination of layer-wise features from a self-supervised model such as Hubert as input to our system. 
Our models only work with image-audio or audio-audio pairs. Extending our models to utilize speech-only data would increase the amount of training data and our framework's applicability.

\bibliographystyle{ieeetr}
\bibliography{ref} 

\begin{thebibliography}{10}

\bibitem{baevski2020wav2vec}
A.~Baevski, Y.~Zhou, A.~Mohamed, and M.~Auli, ``wav2vec 2.0: A framework for
  self-supervised learning of speech representations,'' {\em Advances in neural
  information processing systems}, vol.~33, pp.~12449--12460, 2020.

\bibitem{synnaeve2020end}
G.~Synnaeve, Q.~Xu, J.~Kahn, T.~Likhomanenko, E.~Grave, V.~Pratap, A.~Sriram,
  V.~Liptchinsky, and R.~Collobert, ``End-to-end asr: from supervised to
  semi-supervised learning with modern architectures,'' in {\em ICML 2020
  Workshop on Self-supervision in Audio and Speech}.

\bibitem{wang2020transformer}
Y.~Wang, A.~Mohamed, D.~Le, C.~Liu, A.~Xiao, J.~Mahadeokar, H.~Huang,
  A.~Tjandra, X.~Zhang, F.~Zhang, {\em et~al.}, ``Transformer-based acoustic
  modeling for hybrid speech recognition,'' in {\em ICASSP 2020-2020 IEEE
  International Conference on Acoustics, Speech and Signal Processing
  (ICASSP)}, pp.~6874--6878, IEEE, 2020.

\bibitem{badino2014auto}
L.~Badino, C.~Canevari, L.~Fadiga, and G.~Metta, ``An auto-encoder based
  approach to unsupervised learning of subword units,'' in {\em Acoustics,
  Speech and Signal Processing (ICASSP), 2014 IEEE International Conference
  on}, pp.~7634--7638, IEEE, 2014.

\bibitem{lee2012nonparametric}
C.-y. Lee and J.~Glass, ``A nonparametric bayesian approach to acoustic model
  discovery,'' in {\em Proceedings of the 50th Annual Meeting of the
  Association for Computational Linguistics: Long Papers-Volume 1}, pp.~40--49,
  Association for Computational Linguistics, 2012.

\bibitem{siu2014unsupervised}
M.-h. Siu, H.~Gish, A.~Chan, W.~Belfield, and S.~Lowe, ``Unsupervised training
  of an hmm-based self-organizing unit recognizer with applications to topic
  classification and keyword discovery,'' {\em Computer Speech \& Language},
  vol.~28, no.~1, pp.~210--223, 2014.

\bibitem{kamper2017segmental}
H.~Kamper, A.~Jansen, and S.~Goldwater, ``A segmental framework for
  fully-unsupervised large-vocabulary speech recognition,'' {\em Computer
  Speech \& Language}, vol.~46, pp.~154--174, 2017.

\bibitem{bhati2017unsupervised}
S.~Bhati, S.~Nayak, and K.~S.~R. Murty, ``Unsupervised speech signal to symbol
  transformation for zero resource speech applications,'' {\em Proc.
  Interspeech 2017}, pp.~2133--2137, 2017.

\bibitem{kamper2017embedded}
H.~Kamper, K.~Livescu, and S.~Goldwater, ``An embedded segmental k-means model
  for unsupervised segmentation and clustering of speech,'' in {\em 2017 IEEE
  Automatic Speech Recognition and Understanding Workshop (ASRU)},
  pp.~719--726, IEEE, 2017.

\bibitem{bhati2018phoneme}
S.~Bhati, H.~Kamper, and K.~S.~R. Murty, ``Phoneme based embedded segmental
  k-means for unsupervised term discovery,'' in {\em 2018 IEEE International
  Conference on Acoustics, Speech and Signal Processing (ICASSP)},
  pp.~5169--5173, IEEE, 2018.

\bibitem{oord2018representation}
A.~v.~d. Oord, Y.~Li, and O.~Vinyals, ``Representation learning with
  contrastive predictive coding,'' {\em arXiv preprint arXiv:1807.03748}, 2018.

\bibitem{schneider2019wav2vec}
S.~Schneider, A.~Baevski, R.~Collobert, and M.~Auli, ``wav2vec: Unsupervised
  pre-training for speech recognition,'' {\em Proc. Interspeech 2019},
  pp.~3465--3469, 2019.

\bibitem{chung2021w2v}
Y.-A. Chung, Y.~Zhang, W.~Han, C.-C. Chiu, J.~Qin, R.~Pang, and Y.~Wu,
  ``W2v-bert: Combining contrastive learning and masked language modeling for
  self-supervised speech pre-training,'' in {\em 2021 IEEE Automatic Speech
  Recognition and Understanding Workshop (ASRU)}, pp.~244--250, IEEE, 2021.

\bibitem{bhati21_interspeech}
S.~Bhati, J.~Villalba, P.~Żelasko, L.~Moro-Velázquez, and N.~Dehak,
  ``{Segmental Contrastive Predictive Coding for Unsupervised Word
  Segmentation},'' in {\em Proc. Interspeech 2021}, pp.~366--370, 2021.

\bibitem{bhati2022unsupervised}
S.~Bhati, J.~Villalba, P.~{\.Z}elasko, L.~Moro-Velazquez, and N.~Dehak,
  ``Unsupervised speech segmentation and variable rate representation learning
  using segmental contrastive predictive coding,'' {\em IEEE/ACM Transactions
  on Audio, Speech, and Language Processing}, vol.~30, pp.~2002--2014, 2022.

\bibitem{liu2020mockingjay}
A.~T. Liu, S.-w. Yang, P.-H. Chi, P.-c. Hsu, and H.-y. Lee, ``Mockingjay:
  Unsupervised speech representation learning with deep bidirectional
  transformer encoders,'' in {\em ICASSP 2020-2020 IEEE International
  Conference on Acoustics, Speech and Signal Processing (ICASSP)},
  pp.~6419--6423, IEEE, 2020.

\bibitem{liu2021tera}
A.~T. Liu, S.-W. Li, and H.-y. Lee, ``Tera: Self-supervised learning of
  transformer encoder representation for speech,'' {\em IEEE/ACM Transactions
  on Audio, Speech, and Language Processing}, vol.~29, pp.~2351--2366, 2021.

\bibitem{chen2020uniter}
Y.-C. Chen, L.~Li, L.~Yu, A.~El~Kholy, F.~Ahmed, Z.~Gan, Y.~Cheng, and J.~Liu,
  ``Uniter: Universal image-text representation learning,'' in {\em Computer
  Vision--ECCV 2020: 16th European Conference, Glasgow, UK, August 23--28,
  2020, Proceedings, Part XXX}, pp.~104--120, Springer, 2020.

\bibitem{radford2021learning}
A.~Radford, J.~W. Kim, C.~Hallacy, A.~Ramesh, G.~Goh, S.~Agarwal, G.~Sastry,
  A.~Askell, P.~Mishkin, J.~Clark, {\em et~al.}, ``Learning transferable visual
  models from natural language supervision,'' in {\em International conference
  on machine learning}, pp.~8748--8763, PMLR, 2021.

\bibitem{harwath2018jointly}
D.~Harwath, A.~Recasens, D.~Sur{\'\i}s, G.~Chuang, A.~Torralba, and J.~Glass,
  ``Jointly discovering visual objects and spoken words from raw sensory
  input,'' in {\em Proceedings of the European conference on computer vision
  (ECCV)}, pp.~649--665, 2018.

\bibitem{peng2022fast}
P.~Peng and D.~Harwath, ``Fast-slow transformer for visually grounding
  speech,'' in {\em ICASSP 2022-2022 IEEE International Conference on
  Acoustics, Speech and Signal Processing (ICASSP)}, pp.~7727--7731, IEEE,
  2022.

\bibitem{hsu2019transfer}
W.-N. Hsu, D.~Harwath, and J.~Glass, ``Transfer learning from audio-visual
  grounding to speech recognition,'' {\em Proc. Interspeech 2019},
  pp.~3242--3246, 2019.

\bibitem{peng2022word}
P.~Peng and D.~Harwath, ``Word discovery in visually grounded, self-supervised
  speech models,'' in {\em Interspeech}, 2022.

\bibitem{hsu2021text}
W.-N. Hsu, D.~Harwath, T.~Miller, C.~Song, and J.~Glass, ``Text-free
  image-to-speech synthesis using learned segmental units,'' in {\em
  Proceedings of the 59th Annual Meeting of the Association for Computational
  Linguistics and the 11th International Joint Conference on Natural Language
  Processing}, pp.~5284--5300, 2021.

\bibitem{peng2022fastvgsplus}
P.~Peng and D.~Harwath, ``Self-supervised representation learning for speech
  using visual grounding and masked language modeling,'' in {\em The
  Self-Supervised Learning for Speech and Audio Processing Workshop at AAAI
  2022}, 2022.

\bibitem{wu2022wav2clip}
H.-H. Wu, P.~Seetharaman, K.~Kumar, and J.~P. Bello, ``Wav2clip: Learning
  robust audio representations from clip,'' in {\em ICASSP 2022-2022 IEEE
  International Conference on Acoustics, Speech and Signal Processing
  (ICASSP)}, pp.~4563--4567, IEEE, 2022.

\bibitem{speechclip2022}
Y.-J. Shih, H.-F. Wang, H.-J. Chang, L.~Berry, H.~yi~Lee, and D.~Harwath,
  ``Speechclip: Integrating speech with pre-trained vision and language
  model,'' {\em IEEE SLT}, 2022.

\bibitem{bhati23_interspeech}
S.~Bhati, J.~Villalba, L.~Moro-Velazquez, T.~Thebaud, and N.~Dehak,
  ``{Segmental SpeechCLIP: Utilizing Pretrained Image-text Models for
  Audio-Visual Learning},'' in {\em Proc. INTERSPEECH 2023}, pp.~431--435,
  2023.

\bibitem{nguyen2020zero}
T.~A. Nguyen, M.~de~Seyssel, P.~Roz{\'e}, M.~Rivi{\`e}re, E.~Kharitonov,
  A.~Baevski, E.~Dunbar, and E.~Dupoux, ``The zero resource speech benchmark
  2021: Metrics and baselines for unsupervised spoken language modeling,'' in
  {\em NeuRIPS Workshop on Self-Supervised Learning for Speech and Audio
  Processing}, 2020.

\bibitem{chrupala2022visually}
G.~Chrupa{\l}a, ``Visually grounded models of spoken language: A survey of
  datasets, architectures and evaluation techniques,'' {\em Journal of
  Artificial Intelligence Research}, vol.~73, pp.~673--707, 2022.

\bibitem{synnaeve2014learning}
G.~Synnaeve, M.~Versteegh, and E.~Dupoux, ``Learning words from images and
  speech,'' in {\em NIPS Workshop on Learning Semantics}, Citeseer, 2014.

\bibitem{harwath2016unsupervised}
D.~Harwath, A.~Torralba, and J.~Glass, ``Unsupervised learning of spoken
  language with visual context,'' {\em Advances in Neural Information
  Processing Systems}, vol.~29, 2016.

\bibitem{harwath2017learning}
D.~Harwath and J.~R. Glass, ``Learning word-like units from joint audio-visual
  analysis,'' {\em arXiv preprint arXiv:1701.07481}, 2017.

\bibitem{chrupala2017representations}
G.~Chrupa{\l}a, L.~Gelderloos, and A.~Alishahi, ``Representations of language
  in a model of visually grounded speech signal,'' {\em arXiv preprint
  arXiv:1702.01991}, 2017.

\bibitem{gelderloos2016phonemes}
L.~Gelderloos and G.~Chrupa{\l}a, ``From phonemes to images: levels of
  representation in a recurrent neural model of visually-grounded language
  learning,'' {\em arXiv preprint arXiv:1610.03342}, 2016.

\bibitem{alishahi2017encoding}
A.~Alishahi, M.~Barking, and G.~Chrupa{\l}a, ``Encoding of phonology in a
  recurrent neural model of grounded speech,'' {\em arXiv preprint
  arXiv:1706.03815}, 2017.

\bibitem{harwath2019learning}
D.~Harwath, W.-N. Hsu, and J.~Glass, ``Learning hierarchical discrete
  linguistic units from visually-grounded speech,'' {\em arXiv preprint
  arXiv:1911.09602}, 2019.

\bibitem{guzhov2022audioclip}
A.~Guzhov, F.~Raue, J.~Hees, and A.~Dengel, ``Audioclip: Extending clip to
  image, text and audio,'' in {\em ICASSP 2022-2022 IEEE International
  Conference on Acoustics, Speech and Signal Processing (ICASSP)},
  pp.~976--980, IEEE, 2022.

\bibitem{zhao2021connecting}
Y.~Zhao, J.~Hessel, Y.~Yu, X.~Lu, R.~Zellers, and Y.~Choi, ``Connecting the
  dots between audio and text without parallel data through visual knowledge
  transfer,'' {\em arXiv preprint arXiv:2112.08995}, 2021.

\bibitem{dunbar2021zero}
E.~Dunbar, M.~Bernard, N.~Hamilakis, T.~A. Nguyen, M.~De~Seyssel, P.~Roz{\'e},
  M.~Rivi{\`e}re, E.~Kharitonov, and E.~Dupoux, ``The zero resource speech
  challenge 2021: Spoken language modelling,'' {\em arXiv preprint
  arXiv:2104.14700}, 2021.

\bibitem{hsu2021hubert}
W.-N. Hsu, B.~Bolte, Y.-H.~H. Tsai, K.~Lakhotia, R.~Salakhutdinov, and
  A.~Mohamed, ``Hubert: Self-supervised speech representation learning by
  masked prediction of hidden units,'' {\em IEEE/ACM Transactions on Audio,
  Speech, and Language Processing}, vol.~29, pp.~3451--3460, 2021.

\bibitem{yang21c_interspeech}
S.~wen Yang, P.-H. Chi, Y.-S. Chuang, C.-I.~J. Lai, K.~Lakhotia, Y.~Y. Lin,
  A.~T. Liu, J.~Shi, X.~Chang, G.-T. Lin, T.-H. Huang, W.-C. Tseng, K.~tik Lee,
  D.-R. Liu, Z.~Huang, S.~Dong, S.-W. Li, S.~Watanabe, A.~Mohamed, and
  H.~yi~Lee, ``{SUPERB: Speech Processing Universal PERformance Benchmark},''
  in {\em Proc. Interspeech 2021}, pp.~1194--1198, 2021.

\bibitem{tsai2022superb}
H.-S. Tsai, H.-J. Chang, W.-C. Huang, Z.~Huang, K.~Lakhotia, S.-w. Yang,
  S.~Dong, A.~Liu, C.-I. Lai, J.~Shi, {\em et~al.}, ``Superb-sg: Enhanced
  speech processing universal performance benchmark for semantic and generative
  capabilities,'' in {\em Proceedings of the 60th Annual Meeting of the
  Association for Computational Linguistics}, pp.~8479--8492, 2022.

\bibitem{mpnetv2}
C.~week using JAX/Flax~for NLP and CV, ``Train the best sentence embedding
  model ever with 1b training pairs,'' 2021.

\bibitem{song2020mpnet}
K.~Song, X.~Tan, T.~Qin, J.~Lu, and T.-Y. Liu, ``Mpnet: Masked and permuted
  pre-training for language understanding,'' {\em Advances in Neural
  Information Processing Systems}, vol.~33, pp.~16857--16867, 2020.

\bibitem{wolf-etal-2020-transformers}
T.~Wolf, L.~Debut, V.~Sanh, J.~Chaumond, C.~Delangue, A.~Moi, P.~Cistac,
  T.~Rault, R.~Louf, M.~Funtowicz, J.~Davison, S.~Shleifer, P.~von Platen,
  C.~Ma, Y.~Jernite, J.~Plu, C.~Xu, T.~L. Scao, S.~Gugger, M.~Drame, Q.~Lhoest,
  and A.~M. Rush, ``Transformers: State-of-the-art natural language
  processing,'' in {\em Proceedings of the 2020 Conference on Empirical Methods
  in Natural Language Processing: System Demonstrations}, (Online), pp.~38--45,
  Association for Computational Linguistics, Oct. 2020.

\bibitem{caron2021emerging}
M.~Caron, H.~Touvron, I.~Misra, H.~J{\'e}gou, J.~Mairal, P.~Bojanowski, and
  A.~Joulin, ``Emerging properties in self-supervised vision transformers,'' in
  {\em Proceedings of the IEEE/CVF International Conference on Computer
  Vision}, pp.~9650--9660, 2021.

\end{thebibliography}

\end{document}